\documentclass[amsmath,amssymb,showpacs,twocolumn]{revtex4}
\usepackage{epsfig}
\usepackage{graphicx}
\usepackage{color}

\begin{document}

\title{Entanglement dynamics of two qubits coupled individually to Ohmic baths}

\author{Liwei Duan$^{1}$, Hui Wang$^{2}$, Qing-Hu Chen$^{2}$, Yang Zhao$^{1}$\footnote{Electronic address:~\url{YZhao@ntu.edu.sg}}}
\date{\today}
\address{$^{1}$Division of Materials Science, Nanyang Technological University, Singapore 639798, Singapore \\
$^{2}$Department of Physics, Zhejiang University, Hangzhou, China}

\begin{abstract}
Developed originally for the Holstein polaron, the Davydov D$_1$ ansatz is an efficient, yet extremely accurate trial state for time-dependent
variation of the spin-boson model [J.~Chem.~Phys.~{\bf 138}, 084111 (2013)]. In this work, the Dirac-Frenkel time-dependent variational procedure utilizing the Davydov D$_1$ ansatz is implemented to study entanglement dynamics of two qubits under the influence of two independent baths. The Ohmic spectral density is used without the Born-Markov approximation or the rotating-wave approximation. In the strong coupling regime finite-time disentanglement is always found to exist, while at the intermediate coupling regime, the entanglement dynamics calculated by Davydov D$_1$ ansatz displays oscillatory behavior in addition to entanglement disappearance and revival.
\end{abstract}

\maketitle

\section{Introduction}

Entanglement is an intrinsic property of a quantum system associated with distributed quantum coherence among two or more distinct subsystems.
Beyond its fundamental significance to the theory of quantum mechanics, entanglement also plays an important role in quantum information~\cite{QI1,QI2}, cryptography~\cite{QC}, quantum computation~\cite{QC1,QC2}, quantum teleportation~\cite{QT}, quantum dense coding~\cite{QDC}, and atomic and molecular spectroscopy~\cite{SP1,SP2}.
Due to obvious difficulties to shield a quantum system from environmental decoherence, quantum entanglement is a fragile property under the constant threat of disentanglement. To sustain entanglement in various scenarios is therefore of fundamental importance to quantum information processing~\cite{open,diss}.
Disentanglement in a finite amount of time has been demonstrated experimentally.
For example, Almeida \emph{et al.}~\cite{EPT1} demonstrated the existence of vanishing entanglement by using an all-optical setup, although the environment-induced decay is asymptotic. Laurata \emph{et al.}~\cite{EPT2} studied the behavior of entanglement in atomic ensembles, which includes the dependence of the concurrence on excitation and the quantitative relationship between the local decoherence and entanglement decay.

As early as in 2003, Zhao \emph{et al.}~predicted finite-time entanglement disappearance and revival in a bipartite continuous-variable system \cite{physica}. Yu and Eberly~\cite{YT} found that, for an initially entangled two-qubit system, only a finite time is needed to complete spontaneous disentanglement when the two qubits interact individually with vacuum noise, unlike the local decoherence processe which always takes an infinite time to complete. Ikram \emph{et al.}~\cite{Ikram} showed that disentanglement always exists in the thermal and squeezed reservoirs. Since the Markov approximation was used to deal with the entanglement dynamics in which the bath memory effect is neglected \cite{physica,YT}, much attention was then devoted to the influences of the non-Markovian effects. Bellomo~\emph{et al.}~\cite{Bellomo} proposed a procedure to probe the dynamics of $N$ independent qubits locally coupled to their own baths, which has then been applied to the study of non-Markovian dynamics of two qubits coupled to their individual baths with Lorentzian spectral densities. Entanglement revival (or re-entanglement) has been uncovered despite the absence of direct interactions between the two qubits or their corresponding reservoirs~\cite{Bellomo}, leading to the question of what happens to the lost entanglement when disentanglement occurs. Using the Holstein Hamiltonian, Zhao \emph{et al.}~demonstrated the conversion of exciton-phonon hetero-entanglement to superradiance-like, intraspecies homo-entanglement in 2004 \cite{zz,meier}. L\'{o}pez \emph{et al.}~found in 2008 that the entanglement goes into the bath degrees of freedom at the onset of system disentanglement in a Markovian process, although the two may not happen simultaneously~\cite{Lopez}.
Using a hierarchical equation-of-motion (HEOM) approach, non-Markovian entanglement dynamics has been probed for a two-level system coupled to a common bath in the presence of system-bath coherence by Dijkstra and Tanimura in 2010 \cite{taka}.
An extension to the non-Markovian process has also been reported in Ref.~\cite{ZYXu}, showing a sudden birth of of the bath entanglement as well as entanglement oscillations. Huelga, Rivas and Plenio \cite{Huelga} recently analyzed steady-state entanglement in a two-qubit system coupled to independent non-Markovian baths, in order to capture the influence of Markovian and non-Markovian noises on entanglement preservation. The baths, however, were modeled by an effective noise strength and an adjustable memory time, and therefore, their effect was not treated fully quantum mechanically.

We note that many treatments of the non-Markovian process are based on the rotating wave approximation (RWA) to the spin-boson model while the counter rotating-wave term has been shown recently to play a significant role in the strong coupling regime. Cao~\emph{et al.}~\cite{Cao} proposed an approach based on canonical transformations to deal with the non-Markovian entanglement dynamics of a two-qubit system coupled with individual reservoirs described by the Ohmic spectrum. They found an exponential decay of the entanglement as well as disentanglement in the weak coupling regime. However, the Born approximation is employed in their analysis which means that they can hardly deal with the strong coupling regime, and neither entanglement oscillation or revival was captured in their calculations which is a remarkable signature of the non-Markovian process. Wang and Chen \cite{Wangchen} borrowed the HEOM approach to study the entanglement dynamics of two qubits coupled to Lorentzian baths without using RWA or the Born-Markov approximation. Entanglement sudden death and revival are recovered,
and the counter rotating-wave term has been shown to suppress the entanglement especially in the strong coupling regime.

With great success, the Dirac-Frenkel time-dependent variational approach has been recently employed to treat the spin-boson model using the Davydov D$_1$ ansatz \cite{WN}, a trial state that derives its origin from the Davydov solitons proposed in the 1970s to study transport of biological energy in protein by Davydov and coworkers~\cite{Davydov}. Previously, the Davydov D$_1$ ansatz and its variants had been applied to treat the zero-temperature dynamics of the Holstein molecular crystal model~\cite{Holstein,DH}, while their translationally invariant counterparts were employed to deal with the ground state of the Holstein polaron~\cite{GS}. A two-site form of the Davydov D$_1$ ansatz has been used by Wu \emph{et al.}~\cite{WN} and Chin \emph{et al.}~\cite{Chin} to probe the dynamics and the ground state properties of the sub-Ohmic spin-boson model, respectively. The accuracy of the trial state gains renewed respect after
confirming the existence of non-equilibrium coherent dynamics for strong coupling in the deep sub-Ohmic spin-boson model~\cite{WN,Kast}.
A ground-state trial state was also proposed recently which overcomes the deficiencies of the Silbey-Harris variational method for biased systems \cite{Nazir}.
In this work, the Davydov D$_1$ ansatz is borrowed to study the entanglement dynamics of two qubits coupled with uncorrelated reservoirs, which are taken here as two independent Ohmic spin-boson models.

The rest of paper is structured as follows. In Sec.~II, we introduce the spin-boson Hamiltonian as well as the Davydov D$_1$ ansatz. A two-qubit system coupled to independent bosonic baths is proposed next. In Sec.~III, results on entanglement dynamics calculated by the Davydov D$_1$ ansatz are shown and compared with those under the RWA and in the literature. The conclusions were drawn in Sec.~IV.

\section{Hamiltonian and Ansatz}

\subsection{One qubit}

The  spin-boson model which describes a two-level system (TLS) coupled with a dissipative bosonic bath can be written as
\begin{equation}\label{SB}
\hat{H}_{\rm SB}=-\frac{\Delta}{2}\sigma_x+\sum_l \omega_l b_l^\dag b_l+\frac{\sigma_z}{2}\sum_l \lambda_l(b^\dag_l+b_l),
\end{equation}
where $\hbar=1$ is set to unity, $\sigma_i$ ($i=x, y, z$) are the Pauli matrices, $b_l$ ($b_l^{\dagger}$) is the annihilation (creation) operator of the bosonic bath, $\Delta$ is the tunneling amplitude of the TLS, $\omega_l$ is the frequency of mode $l$, and $\lambda_l$ is the corresponding coupling strength determined by the spectral density $J(\omega)$. In this paper, the spectral density $J(\omega)$ is given by
\begin{eqnarray}\label{OspectraZ}
J(\omega)=\sum_l \lambda^2_l \delta(\omega-\omega_l)=2\alpha\omega_c^{1-s}\omega^s \Theta(\omega-\omega_c),
\end{eqnarray}
where $\alpha$ is a dimensionless coupling constant, $\omega_c$ is the cutoff frequency, $\Theta(\omega-\omega_c)$ is a step function, and $s$ is the spectral exponent. Depending on the value of spectral exponent $s$, there exist three types of the bath spectral density~\cite{Leggett}:
sub-Ohmic ($s<1$), Ohmic ($s=1$), and super-Ohmic ($s>1$).

The Ohmic regime is sandwiched between the sub-Ohmic regime with a delocalization-localization phase transition~\cite{Chin,Zhang,Bulla,Alvermann} and the super-Ohmic regime with no phase transition~\cite{Leggett}. The spin-boson model can be mapped onto the anisotropic Kondo model by using bosonization techniques, and it is well understood that there exists a Kosterlitz-Thouless-type phase transition for the Ohmic bath~\cite{Leggett,diss}.
As mentioned earlier,
the HEOM approach is applicable to entanglement dynamics under influence of Lorentzian bath spectral densities~\cite{Wangchen}. It has also been extended to study long-lasting electronic coherence in the presence of
a superposition of shifted Lorentzian modes by Kreisbeck and Kramer~\cite{SL}, and to probe
electron transfer dynamics with a Brownian oscillator spectral density by Tanaka and Tanimura~\cite{TT}.
To the best of our knowledge, however, the HEOM method has not been employed to deal with entanglement dynamics in the Ohmic regime, where we hope our variational approach can shed light on the bath effect on disentanglement.
Therefore, in this work we
consider only the Ohmic bath, which has been used to describe a wide range of physical and chemical processes, such as the electron-hole drag of charged particles in metals, the quasiparticle tunneling in Josephson junctions~\cite{diss}, and exciton transport in pigment-protein complexes under the influences of solvent environments and protein fluctuations~\cite{Gilmore}.

The Hamiltonian $\hat{H}_{\rm SB}$ has been analyzed by the Davydov D$_1$ ansatz which has the form
\begin{eqnarray}
|\phi(t)\rangle&=&A(t)|+\rangle\exp[\sum_l (f_l(t)b_l^\dag-h.c.)]|0\rangle_{\rm B}\nonumber\\
\label{trial func}
&+&B(t)|-\rangle\exp[\sum_l (g_l(t)b_l^\dag-h.c.)]|0\rangle_{\rm B},
\end{eqnarray}
where $|+\rangle$ ($|-\rangle$) is the spin up (down) state and $|0\rangle_{\rm B}$ is the vacuum state of the bosonic bath.
$A(t)$ and $B(t)$ are variational parameters representing occupation amplitudes in states $|+\rangle$ and $|-\rangle$, respectively, and $f_l(t)$ and $g_l(t)$ ($l=1, 2, 3, ...$) label the corresponding  phonon displacements of mode $l$. Detailed procedures on how to get the equations of motion for the variational parameters can be found in the appendix.

\begin{table*}[tbhp]
\centering
\caption{The detailed transformations for the Hamiltonian, wave function and density matrix.}
\begin{tabular}{p{100pt}p{100pt}p{100pt}}
  \hline\hline
   & $\hat{H}_{\rm SB}$ & $\hat{H}_{\rm RSB}$ \\
  \hline
  Hamiltonian & $\hat{H}_{\rm SB}=U\hat{H}_{\rm RSB}U^{\dagger}$ & $\hat{H}_{\rm RSB}=U^{\dagger}\hat{H}_{\rm SB}U$ \\
  Wave function & $|\psi(t)\rangle=U|\phi(t)\rangle$ & $|\phi(t)\rangle=U^{\dagger}|\psi(t)\rangle$  \\
    Density matrix & $\rho_{\rm SB}=U\rho_{\rm RSB}U^{\dagger}$ & $\rho_{\rm RSB}=U^{\dagger}\rho_{\rm SB} U$ \\
  \hline\hline
\end{tabular}
\label{trans}
\end{table*}

It should be noted that generally the  spin-boson model is not directly used to the entanglement calculations. In previous entanglement studies, the Hamiltonian used can be transformed to the  spin-boson model by a $90^{\circ}$ rotation, and vice versa, by a rotating operator $U=\exp{(\frac{i\pi}{4}\sigma_y)}$:
\begin{eqnarray}\label{RSB}
\hat{H}_{\rm RSB}&=&U^{\dagger}\hat{H}_{\rm SB}U\nonumber\\
&=&\frac{\Delta}{2}\sigma_z+\sum_l \omega_l b_l^\dag b_l+\frac{\sigma_x}{2}\sum_l \lambda_l(b^\dag_l+b_l).
\end{eqnarray}
If RWA is considered, the spin-boson interaction $\frac{\sigma_x}{2}\sum_l \lambda_l(b^\dag_l+b_l)$ can be simplified as $\frac{1}{2}\sum_l \lambda_l(b^\dag_l\sigma_- + b_l\sigma_+)$, where $\sigma_{\pm}=\frac{1}{2}(\sigma_x \pm i\sigma_y)$. For the Hamiltonian with RWA,  the total excitation number is conserved, since $\hat{N}=\sigma_+ \sigma_- + \sum_l b^{\dagger}_l b_l$ commutes with the Hamiltonian~(\ref{RSB}) under the RWA.
As we are interested in the dynamics in $\hat{H}_{\rm RSB}$, we can transform the Davydov D$_1$ ansatz in $\hat{H}_{\rm SB}$ into $|\psi(t)\rangle=U^{\dagger}|\phi(t)\rangle$. Alternatively, we can calculate the dynamics in $\hat{H}_{\rm SB}$, and transform the final results into $\hat{H}_{\rm RSB}$. Detailed properties transformations  are shown in Table~\ref{trans}.

For comparisons, we also calculated the exact results under the RWA. Since the total excitation number is conserved, we  focus on the certain excitation number subspace depending on the initial conditions. Here we consider the zero- and one-excitation subspace in which the complete bases can be written as $|-\rangle|0\rangle_{\rm B}$, $|+\rangle|0\rangle_{\rm B}$ and $|-\rangle b^{\dagger}_l|0\rangle_{\rm B}$. The time-dependent wave function can be written as:
\begin{eqnarray}
|\varphi(t)\rangle=c |-\rangle|0\rangle_{\rm B} + c_0(t) |+\rangle|0\rangle_{\rm B} + \sum_l c_l(t) |-\rangle b^{\dagger}_l|0\rangle_{\rm B}.
\end{eqnarray}
From the time-dependent Schr\"{o}dinger equation, we can derive a set of self-consistent equations for the amplitudes $c_0(t)$ and $c_l(t)$. Then we can get the wave function as well as the reduced density matrix. The detailed derivations can be found in Ref.~\cite{open}.

\subsection{Two qubits}

In order to study the entanglement dynamics, two independent spin-boson models need to be considered simultaneously. The total Hamiltonian can be written as
\begin{eqnarray}
\hat{H}_{\rm T}=\hat{H}_1+\hat{H}_2,
\end{eqnarray}
where $\hat{H}_{\rm i}$ $(i=1,2)$ is  $\hat{H}_{\rm RSB}$ for qubit $i$. The anti-Bell state and the Bell state are chosen in this work as the initial state:
\begin{eqnarray}
|\Psi(0)\rangle_{\rm anti}&=&a|+-\rangle+\sqrt{1-a^2}|-+\rangle,\label{anti}\\
|\Psi(0)\rangle_{\rm Bell}&=&a|++\rangle+\sqrt{1-a^2}|--\rangle.\label{bellstate}
\end{eqnarray}
Take the anti-Bell initial state as an example. As the two spin-boson Hamiltonians are completely decoupling, they evolve independently.
The time-dependent wave functions $|\psi_i(t)\rangle$ for qubit $i$ ($i=1,2$) can be readily obtained from the variational procedure, and the total wave function of the two-qubit system takes the form
\begin{eqnarray}
|\Psi(t)\rangle&=&e^{-i\hat{H}t}|\Psi(0)\rangle\nonumber\\
&=&e^{-i(\hat{H}_1+\hat{H}_2)t}(a|+\rangle_1|-\rangle_2
+\sqrt{1-a^2}|-\rangle_1|+\rangle_2)\nonumber\\
&=&a e^{-i\hat{H}_1 t}|+\rangle_1e^{-i\hat{H}_2 t}|-\rangle_2\nonumber\\
&&+\sqrt{1-a^2} e^{-i\hat{H}_1 t}|-\rangle_1 e^{-i\hat{H}_2 t}|+\rangle_2\nonumber\\
&=&a |\psi_1(t)\rangle_+ |\psi_2(t)\rangle_-\nonumber\\
&&+\sqrt{1-a^2} |\psi_1(t)\rangle_- |\psi_2(t)\rangle_+,
\end{eqnarray}
where the subscript $+ (-)$ in $|\psi_i(t)\rangle_{\pm}$ indicates whether the qubit is initially in the spin up($+$) or down($-$) state, and $\exp{(-i\hat{H}_i t)}|\pm\rangle_i$ has been replaced by $|\psi_i(t)\rangle_{\pm}$ obtained from the variational procedure. For example,
$|\psi_i(0)\rangle_{+}$ can be obtained by setting, for qubit $i$, the initial variational parameters: $A(0)=1$, $B(0)=0$, and $f_l(0)=g_l(0)=0$. The time evolution of
$|\psi_i(t)\rangle_{+}$ is determined by the Dirac-Frenkel time-dependent variation.

 After obtaining the total wave function of the two-qubit system, the reduced density matrix $\rho^T(t)$ can be evaluated by
\begin{eqnarray}
\rho^T(t)={\rm Tr}_{{\rm B}_1,{\rm B}_2}( |\Psi(t)\rangle\langle\Psi(t)|).
\end{eqnarray}
Here we choose the two-qubit product states $|1\rangle=|++\rangle$, $|2\rangle=|+-\rangle$, $|3\rangle=|-+\rangle$, and $|4\rangle=|--\rangle$ as the bases.

Entanglement between the qubits is quantified in this work by the concurrence~\cite{Wootters}, which is defined as
\begin{equation}\label{con}
C=\max{(0,\sqrt{e_1}-\sqrt{e_2}-\sqrt{e_3}-\sqrt{e_4})},
\end{equation}
where $e_i$ are the eigenvalues of the matrix $\rho^T (\sigma_y^1\otimes\sigma_y^2)\rho^{T*} (\sigma_y^1\otimes\sigma_y^2)$, indexed in a decreasing order. Concurrence $C$ varies from $0$ for a completely disentangled state to $1$ for a maximally entangled state. For $\hat{H}_{\rm RSB}$ and an initial anti-Bell or Bell state, the density matrix takes the form
\begin{equation}
\rho^T=\begin{pmatrix}
\rho^T_{11} & 0 & 0 & \rho^T_{14} \\
0 & \rho^T_{22} & \rho^T_{23} & 0 \\
0 & \rho^T_{32} & \rho^T_{33} &0 \\
\rho^T_{41} & 0 & 0 & \rho^T_{44}
\end{pmatrix}.
\end{equation}
The concurrence in this case can be simplified as~\cite{Ikram,Bellomo}
\begin{eqnarray}
C=2\max{\left(0,|\rho^T_{14}|-\sqrt{\rho^T_{22}\rho^T_{33}}, |\rho^T_{23}|-\sqrt{\rho^T_{11}\rho^T_{44}}\right)}.
\end{eqnarray}

\section{Results and Discussions}

\subsection{Concurrence for $\hat{H}_{\rm RSB}$}

\begin{figure}[t]
\includegraphics[scale=0.88]{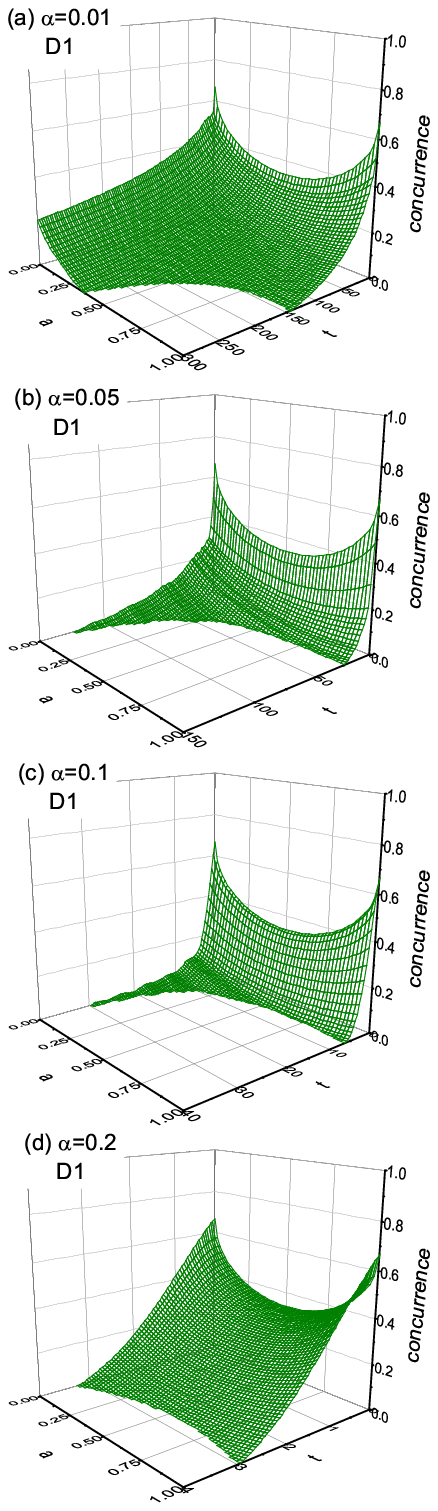}
\includegraphics[scale=0.88]{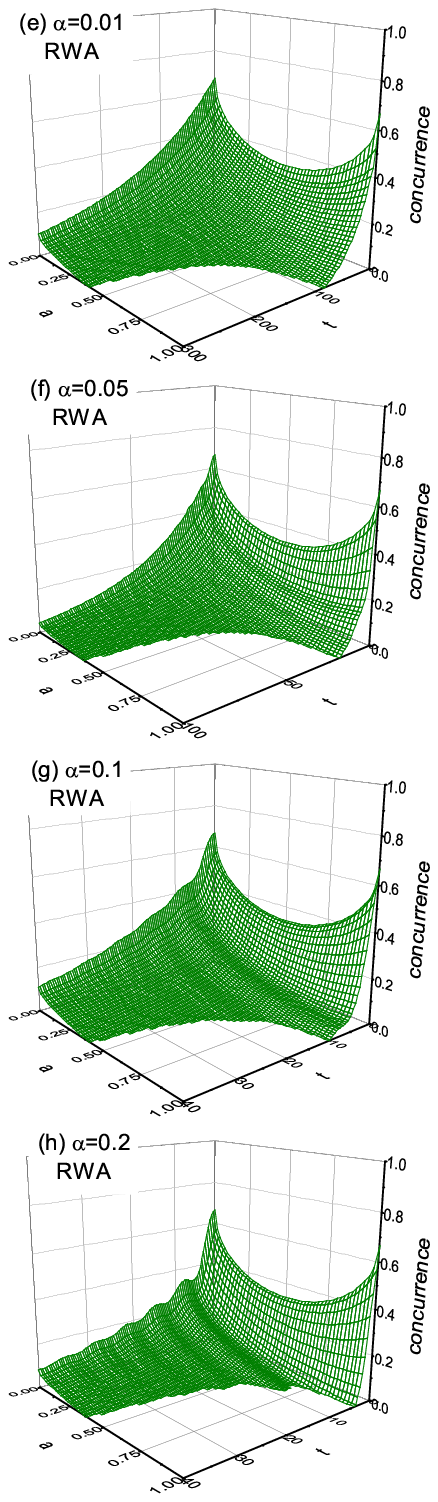}
\caption{Time evolution of the concurrence as a function of $a$ calculated by the Davydov D$_1$ ansatz [(a)-(d)] and under the RWA [(e)-(h)]. The two qubits are initially in the mixed state characterized by $a$. $\Delta=0.2$. (a) and (e): $\alpha=0.01$; (b) and (f): $\alpha=0.05$; (c) and (g): $\alpha=0.1$; (d) and (h): $\alpha=0.2$.}
\label{zheng}
\end{figure}

The concurrence for $\hat{H}_{\rm RSB}$ has received much attention in the literature. Most of the studies focus on the Lorentzian bath spectral density, for which there is an obvious distinction between the Markovian and non-Markovian regimes, and exact results are known to exist under the RWA. Cao \emph{et al.}~studied the effect of the Ohmic spectra using an approach based on the polaron canonical transformation, and their chosen initial state is a mixed one given by \cite{vedral,YT,Cao}
\begin{eqnarray}\label{ini}
\rho^T(0)&=&\frac 1 3 (1-a)|--\rangle\langle--| +\frac a 3|++\rangle\langle++|\nonumber\\ &+&\frac 1 3(|+-\rangle+|-+\rangle)(\langle+-|+\langle-+|).
\end{eqnarray}
Fig.~\ref{zheng} compares the entanglement dynamics as a function of $a$ and $t$
calculated by D$_1$ ansatz with that evaluated exactly under the RWA
for various values of the coupling strength $\alpha$.
The same sets of parameters from Ref.~\cite{Cao} are chosen in Figs.~\ref{zheng} (a)-(c) and (e)-(g) to facilitate the comparison. In Figs.~\ref{zheng} (d) and (h),
we also calculate entanglement dynamics for a strong-coupling case, which is absent in Ref.~\cite{Cao} due to the perturbative nature of their approach.

As shown in Figs.~\ref{zheng} (a)-(c) and (e)-(g), in the weak coupling regime, the entanglement relaxation properties calculated by both approaches show considerable changes as one varies $a$ in the initial state of Eq.~(\ref{ini}) despite starting from similar values of the initial concurrence. By decreasing $a$, therefore reducing the portion of the doubly-excited component in the initial density matrix, the entanglement lifetime can be prolonged. In the strong coupling regime,
however, the differences become less obvious, as shown in Fig.~\ref{zheng} (d). These observations can be explained by the competition among various Hamiltonian terms in Eq.~(\ref{RSB}).
In the weak coupling regime,
the spin Hamiltonian ${\Delta}\sigma_z /2$ dominates the time evolution, and the zero excitation state $|--\rangle$ is the ground state of the two-qubit system with a much lower energy. Decreasing $a$, therefore increasing the content of the zero-excitation component in the initial density matrix, the concurrence will last for a longer time. Similar conclusions have also been drawn in Ref.~\cite{Ikram}. In the strong coupling regime, the interaction Hamiltonian in Eq.~(\ref{RSB}), $\sigma_x/2\sum_l\lambda_l(b_l^{\dagger}+b_l)$, instead plays a dominant role. Although the zero excitation state may still have a lower energy, the difference all but vanishes, leading to independence of the entanglement lifetime on $a$.

If we compare the left and right columns in Fig.~\ref{zheng}, we are convinced that the the two methods employed in this work generate quite similar entanglement dynamics
in the weak coupling regime, where the Hamiltonian under the RWA is regarded as a good approximation to  $\hat{H}_{\rm RSB}$.
The agreement is especially good in Figs.~\ref{zheng} (a) and (e), which also indicates that the Davydov D$_1$ ansatz is robust in the weak coupling regime.
However, differences between the two approaches emerge upon entering the strong coupling regime where the RWA entanglement dynamics for small $a$ is found
to survive for a much longer time than that calculated by the Davydov D$_1$ ansatz.
The reduced the entanglement  lifetime obtained from the D$_1$ ansatz as compared to the RWA counterpart confirmed
that the counter rotating-wave term suppresses the inter-qubit entanglement, especially in the strong coupling regime, a conclusion reached in Ref.~\cite{Wangchen}.
Comparisons also show that the entanglement dynamics derived from the D$_1$ ansatz is consistent with that in Ref.~\cite{Cao}, but differences exist in the strong coupling regime.

Another interesting feature in the results derived from the D$_1$ ansatz is the small-amplitude oscillations in the concurrence that emerge in the intermediate coupling regime, as shown in Figs.~\ref{zheng} (b) and (c). Under the RWA, such oscillations are also found to exist,
but they are completely absent from the perturbative results in Ref.~\cite{Cao}.
As is known, the oscillating behavior of the entanglement is a signature of the bath non-Markovian effect. In Ref.~\cite{Cao},
the entanglement dynamics only displays a monotonic decay. It is difficult to be certain on whether the non-Markovian effect was included in Ref.~\cite{Cao}, since the monotonic decay of the entanglement can be induced by the Markovian effect as well.
The use of the Born approximation and the omission of higher order excitations in the transformed Hamiltonian renders the approach in Ref.~\cite{Cao} less reliable as the coupling strength is increased. On the other hand, the Davydov D$_1$ ansatz has been known to become more accurate with the increase of the coupling strength~\cite{WN}, and therefore, we expect our time-dependent variation to be robust in the intermediate coupling regime.

\begin{figure}[htbp]
\centering
\includegraphics[scale=0.66]{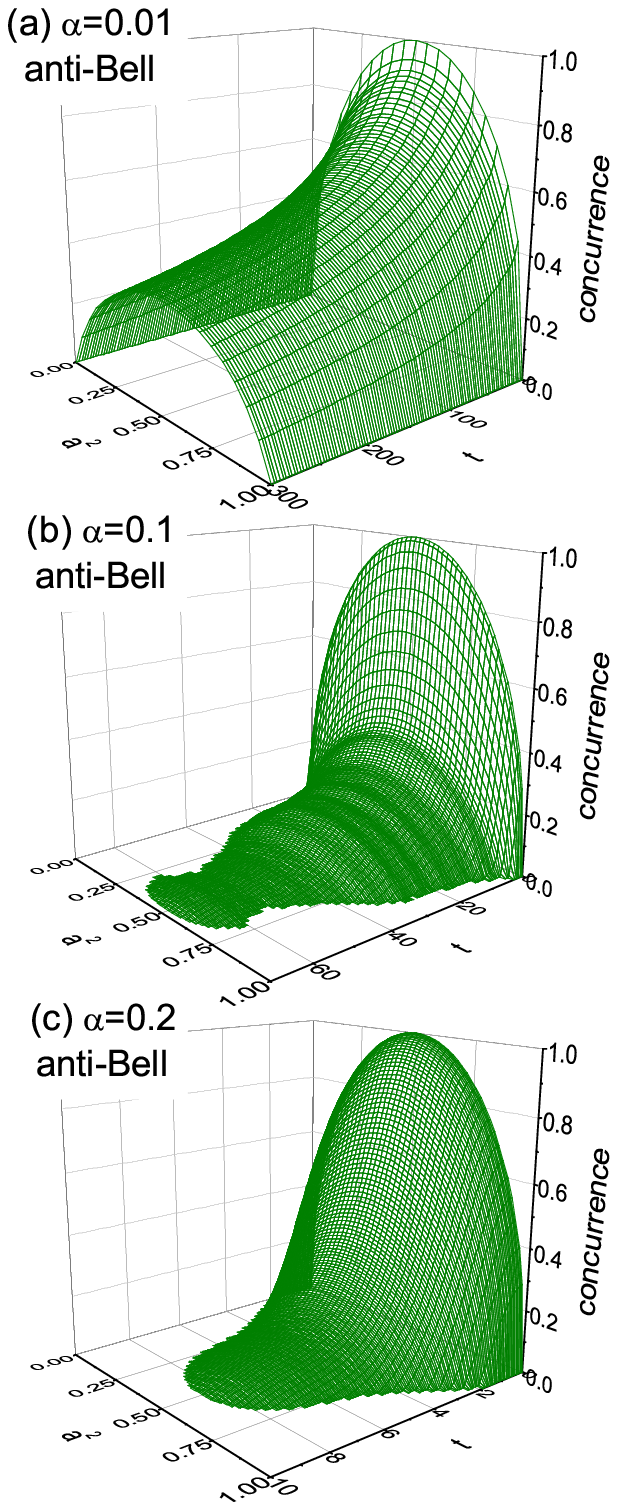}
\includegraphics[scale=0.66]{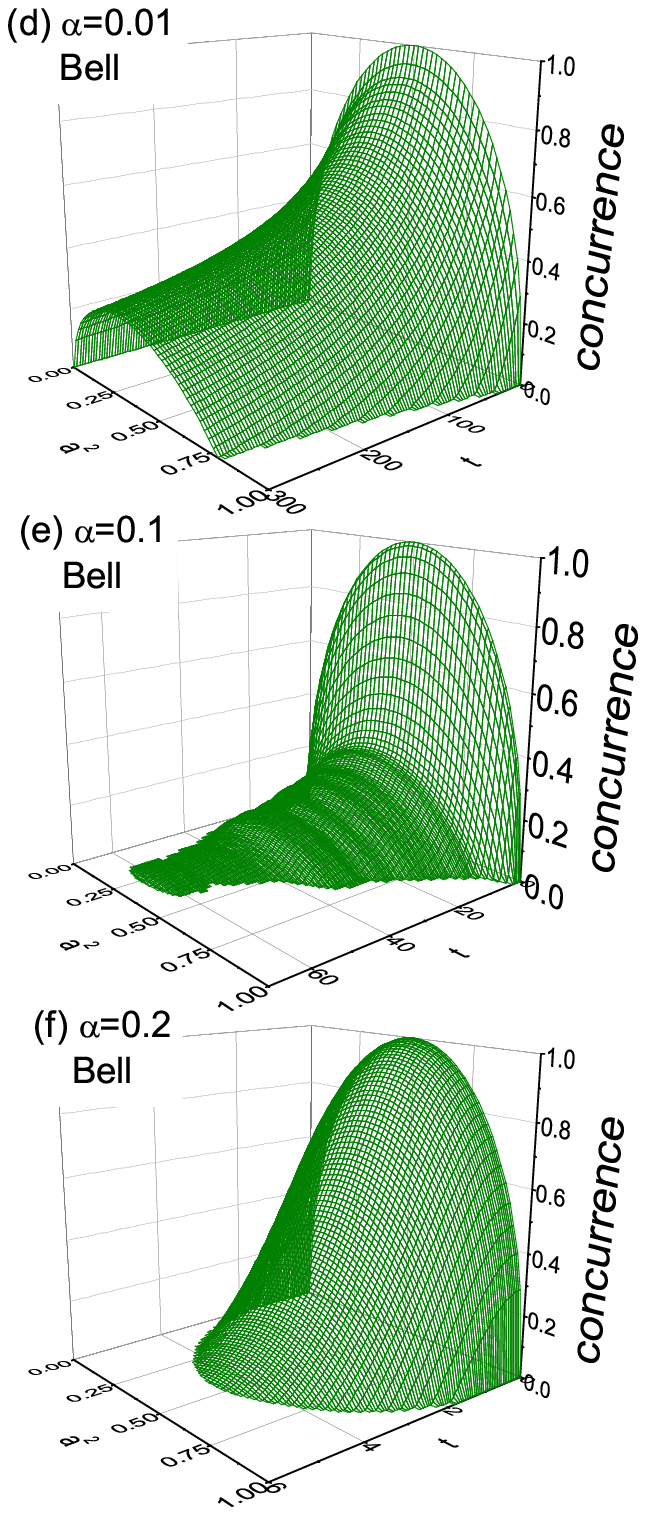}
\caption{Time evolution of the concurrence as a function of $a^2$ calculated by the Davydov D$_1$ ansatz. The two qubits are initially in the anti-Bell state (left panel) and Bell state (right panel). $\Delta=0.2$. (a) and (d): $\alpha=0.01$; (b) and (e): $\alpha=0.1$; (c) and (f): $\alpha=0.2$.}
\label{zheng-anti}
\end{figure}

In search for more pronounced oscillating behavior in the entanglement dynamics,
the anti-Bell and Bell states are next employed as the initial states, and results calculated by the D$_1$ ansatz are displayed in Fig.~\ref{zheng-anti} as a function of $a^2$ and $t$. It is found that the concurrence in Fig.~\ref{zheng-anti} lasts for a much longer time than that calculated previously for an initial mixed state.
The entanglement dynamics for an initial
anti-Bell state is symmetric about $a^2=0.5$, while that for an initial Bell state display no symmetry.
This is understood as follows.
From anti-Bell states, the total wave function of the two-qubit system is always symmetric under the exchange of the two spin-boson systems considered, while such a symmetry in the Bell state is only preserved when the population is equally shared by the two parties during the time evolution.

It is interesting to observe, for the first time, concurrence oscillations, disentanglement and re-entanglement in a single plot of the concurrence time evolution for  $\alpha=0.1$ in Figs.~\ref{zheng-anti} (b) and (e), for initial correlated and anti-correlated Bell states, respectively. To the best of our knowledge, no such observations were reported for the Ohmic bath. A question naturally arises: why do the concurrence oscillations, disentanglement and re-entanglement only exist in the intermediate coupling regime? As shown in the work of Wang and Chen \cite{Wangchen}, the non-Markovian effect is always found to be suppressed by strong spin-bath coupling.
On the one hand, the non-Markovian effect feeds back information to the system with adjustments to the atomic state, and the entanglement is strengthened which leads to revival. On the other hand, the strong spin-bath coupling immobilizes the atomic states, making it harder to recover the vanished entanglement. When the coupling is sufficiently strong, entanglement revival ceases completely with the disappearance of the non-Markovian effect.

\subsection{Concurrence for $\hat{H}_{\rm SB}$}

\begin{figure}[htbp]
\centering
\includegraphics[scale=0.46]{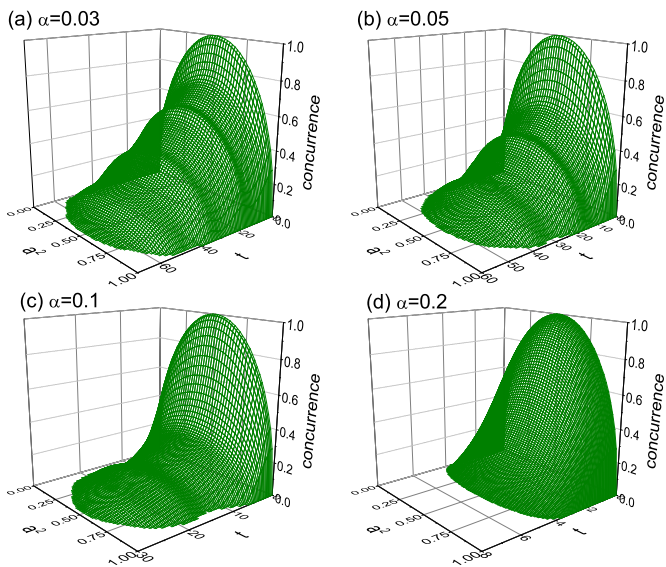}
\caption{Time evolution of the concurrence as a function of $a$ when the two qubits are initially in the anti-Bell state for $\Delta=0.2$. (a) $\alpha=0.03$, (b) $\alpha=0.05$, (c) $\alpha=0.1$, and (d) $\alpha=0.2$.}
\label{original}
\end{figure}

The entanglement dynamics for $\hat{H}_{\rm RSB}$ has received much attention, but few studies have been carried out on the entanglement dynamics of the original Hamiltonian before the rotation, $\hat{H}_{\rm SB}$. As the concurrence given by Eq.~(\ref{con}) only depends on the eigenvalues, the application of
the unitary transformation $U$ shall not change the value of the concurrence.
For example, if the Bell state, Eq.~(\ref{bellstate}), is taken as the initial state for $\hat{H}_{\rm RSB}$, the concurrence will be the same if,
for $\hat{H}_{\rm SB}$, the following transformed initial state is used:
\begin{equation}
U|\Psi(0)\rangle_{\rm Bell}=a|--\rangle_x+\sqrt{1-a^2}|++\rangle_x
\end{equation}
where $|\pm\rangle_x$ are the eigenstates of $\sigma_x$ with eigenvalues 1 and -1, respectively.
However, if the original initial state $|\Psi(0)\rangle_{\rm Bell}$ is used for $\hat{H}_{\rm SB}$, which is equivalent to taking $U^{\dagger}|\Psi(0)\rangle_{\rm Bell}$ as the initial state for $\hat{H}_{\rm RSB}$, the concurrence so calculated will no longer remain the same.

It was known previously through variational studies of the Holstein polaron that the accuracy of the Davydov D$_1$ ansatz increases with an increasing exciton-phonon coupling strength. Some recent comparisons of the variational results with those from the adaptive time-dependent density matrix renormalization group~\cite{yao} have demonstrated an unexpectedly robust precision of the Davydov D$_1$ ansatz even in the weak coupling regime.
In Fig.~\ref{original}, we show entanglement dynamics of $\hat{H}_{\rm SB}$ as calculated by the Davydov D$_1$ ansatz.
The anti-Bell state, Eq.~(\ref{anti}), is chosen as the initial state.
Due to the symmetry in Hamiltonian~(\ref{SB}), the two types of initial conditions, the Bell state and the anti-Bell state, actually lead to the same concurrence dynamics, an outcome at variance with the case of $\hat{H}_{\rm RSB}$. As shown in Fig.~\ref{original}, there is an additional symmetry axis of $a^2=0.5$ in the entanglement dynamics of $\hat{H}_{\rm SB}$.
Oscillations of the concurrence as a function of time are easily seen in the intermediate coupling regime, and an increase in the coupling strength will lead to entanglement suppression which ends with an early disappearance of the concurrence for strong coupling.

\section{Conclusions}

The entanglement dynamics for a system of two qubits coupled individually to Ohmic bosonic baths have been studied by employing the powerful, numerically efficient Davydov D$_1$ ansatz.
Exact results for the Hamiltonian under the RWA have also been calculated as comparisons.
In the weak coupling regime, RWA is a quite good approximation to the full spin-boson Hamiltonian, and these two methods show quite good agreement lending support to the  validity of the Davydov D$_1$ ansatz.

Significate influences over the entanglement dynamics are found to come from the initial portion of the doubly-excited state, the increase of which is shown to reduce considerably the entanglement lifetime. Unable to describe the high order excitations induced by the counter rotating-wave term as coupling increases, the RWA loses its validity in the strong coupling regime. But the Davydov D$_1$ ansatz becomes more accurate as has been discussed in Ref.~\cite{WN}. The initial portion of the doubly-excited state has only minor influences over the entanglement dynamics of the Hamiltonian without RWA in the strong coupling regime.
We also compared our results with those in Ref.~\cite{Cao}, finding agreement in the weak coupling regime.
Due to the Born approximation used in Ref.~\cite{Cao}, the method is incapable to deal with the strong coupling regime.

Among our findings in the intermediate coupling regime are the oscillating behavior of the entanglement and its revival,
which are typical signatures of non-Markovian processes,
for both the Bell and anti-Bell initial states using an Ohmic bath spectral density.
The disentanglement is found to always exist in the strong coupling regime, and
the entanglement lifetime calculated by the Davydov D$_1$ ansatz is shortened substantially
as compared with that under the RWA, confirming the entanglement-suppressing effect of the counter rotating-wave term previously discovered in Ref.~\cite{Wangchen}.

The concurrence employed here does not possess an operational interpretation despite myriad studies of it as an entanglement measure. The mixed-state concurrence can be obtained as a convex roof \cite{Wootters,conv}, but it has been demonstrated by Plenio \cite{convex} that the convexity is merely a mathematical requirement for entanglement monotones, and generally does not correspond to a physical process describing the loss of information about a quantum system. In contrast, the logarithmic negativity is an entanglement monotone under with an operational interpretation \cite{convex,LN}. Study based on the logarithmic negativity of a two-qubit system can be an interesting future direction.

\vspace{1cm}

\section*{Acknowledgements}
Support from the Singapore National Research Foundation through the Competitive Research Programme (CRP)
under Project No.~NRF-CRP5-2009-04 is gratefully acknowledged. This work is also supported in part
by the National Natural Science Foundation of
China under Grant No~11174254.

\appendix*
\section{The variational procedure}
In this work, the Lagrangian formalism of the Dirac-Frenkel variational principle is employed to obtain the equations of motion for the variational parameters~\cite{WN}. The Lagrangian operator is defined as
\begin{eqnarray}\label{lag}
L&=&\langle D_1(t)|\frac{i}{2}\frac{\overleftrightarrow{\partial}}{\partial t}-\hat{H}|D_1(t)\rangle\nonumber\\
&=&\frac{i}{2}\left[\langle D_1(t)|\frac{\overrightarrow{\partial}}{\partial t}|D_1(t)\rangle-\langle D_1(t)|\frac{\overleftarrow{\partial}}{\partial t}|D_1(t)\rangle\right]\nonumber\\
&&-\langle D_1(t)\hat{H}|D_1(t)\rangle~.
\end{eqnarray}

We can get the equation of motion for the variational parameter $u_n$ from
\begin{equation}\label{lag eq}
\frac{d}{dt}\left(\frac{\partial L}{\partial \dot{u}_n^*}\right)-\frac{\partial L}{\partial u_n^*}=0,
\end{equation}
where $u_n^*$ denotes the complex conjugate of  $u_n$, which can be $A$, $B$, $f_l$ or $g_l$.
From Eq.~(\ref{lag eq}).
The equations of motion for $A(t)$ and $B(t)$ can be written as
\begin{eqnarray}
0=&&i\dot{A}+i\frac{A}{2}\sum_l(\dot{f}_lf_l^*-\dot{f}_l^*f_l)-\frac{A}{2}\sum_l\lambda_l (f_l+f_l^*)\nonumber\\
\label{eq A}
&+&\frac{\Delta B}{2}e^{\sum_l \left[f_l^*g_l-\frac{1}{2}(|f_l|^2+|g_l|^2)\right]}-A\sum_l\omega_l|f_l|^2~,\\
0=&&i\dot{B}+i\frac{B}{2}\sum_l(\dot{g}_lg_l^*-\dot{g}_l^*g_l)+\frac{B}{2}\sum_l\lambda_l (g_l+g_l^*)\nonumber\\
\label{eq B}
&+&\frac{\Delta A}{2}e^{\sum_l \left[g_l^*f_l-\frac{1}{2}(|f_l|^2+|g_l|^2)\right]}-B\sum_l\omega_l|g_l|^2~.
\end{eqnarray}
Similarly, the equations of motion for $f_l(t)$ and $g_l(t)$ can be written as
\begin{eqnarray}
&&iA\dot{f}_l-A\frac{\lambda_l}{2}-A\omega_l f_l\nonumber\\
\label{eq f}
&=&\frac{\Delta B}{2}(f_l-g_l)e^{\sum_l \left[f_l^*g_l-\frac{1}{2}(|f_l|^2+|g_l|^2)\right]}~,\\
&&iB\dot{g}_l+B\frac{\lambda_l}{2}-B\omega_l g_l\nonumber\\
\label{eq g}
&=&\frac{\Delta A}{2}(g_l-f_l)e^{\sum_l \left[g_l^*f_l-\frac{1}{2}(|f_l|^2+|g_l|^2)\right]}~.
\end{eqnarray}
The time-dependent wave function and the density matrix can be calculated accordingly after we solve the equations of motion.


%
%
%

\end{document}